\documentclass[%
 aip,
 amsmath,amssymb,
 reprint,%
]{revtex4-1}
\usepackage{subfigure}
\usepackage{hyperref}
\usepackage{booktabs}

\usepackage{graphicx}
\usepackage{dcolumn}
\usepackage{cleveref}
\usepackage[abbreviations]{glossaries-extra}

\usepackage[utf8]{inputenc}
\usepackage[T1]{fontenc}
\usepackage{threeparttable}
\usepackage{mathptmx}
\usepackage{etoolbox}
\usepackage{color}
\usepackage{CJKutf8}
\usepackage{xcolor}

\usepackage{soul}

\soulregister\cite7
\soulregister\ref7
\soulregister\eqref7
\soulregister\pageref7 

\makeatletter
\def\@email#1#2{%
 \endgroup
 \patchcmd{\titleblock@produce}
  {\frontmatter@RRAPformat}
{\frontmatter@RRAPformat{\produce@RRAP{*#1\href{mailto:#2}{#2}}}\frontmatter@RRAPformat}
  {}{}
}%
\makeatother
\begin{document}

\preprint{AIP/123-QED}

\title[]{Comparative study of flow fluctuations in ruptured and unruptured intracranial aneurysms: A lattice Boltzmann study}
\author{F.~Huang}
\email{seyed.hosseini@ovgu.de.}
\affiliation{Laboratory of Fluid Dynamics and Technical Flows, University of Magdeburg ``Otto von Guericke'', D-39106 Magdeburg, Germany.}%
\author{S.A.~Hosseini}%
\affiliation{Laboratory of Fluid Dynamics and Technical Flows, University of Magdeburg ``Otto von Guericke'', D-39106 Magdeburg, Germany.}%
\affiliation{Department of Mechanical and Process Engineering, ETH Z\"urich, 8092 Z\"urich, Switzerland.}%
\author{G.~Janiga}
\affiliation{Laboratory of Fluid Dynamics and Technical Flows, University of Magdeburg ``Otto von Guericke'', D-39106 Magdeburg, Germany.}%
\author{D.~Th\'evenin}
\affiliation{Laboratory of Fluid Dynamics and Technical Flows, University of Magdeburg ``Otto von Guericke'', D-39106 Magdeburg, Germany.}%

\newabbreviation{cfd}{CFD}{computational fluid dynamics}
\newabbreviation{piv}{PIV}{particle image velocimetry}
\newabbreviation{chmrt}{CHMRT}{central Hermite multiple relaxation time}
\newabbreviation{srt}{SRT}{single relaxation time}
\newabbreviation{lbm}{LBM}{lattice Boltzmann method}
\newabbreviation{dns}{DNS}{direct numerical simulation}
\newabbreviation{lr}{LR}{Low resolution}
\newabbreviation{mr}{MR}{Middle resolution}
\newabbreviation{hr}{HR}{High resolution}
\newabbreviation{cfl}{CFL}{Courant-
Friedrichs-Lewy}
\newabbreviation{fke}{FKE}{Fluctuating (or turbulent) kinetic energy}
\newabbreviation{psd}{PSD}{power spectral density}
\newabbreviation{fft}{FFT}{fast Fourier transform}

\date{\today}
\begin{abstract}
Flow fluctuations have recently emerged as a promising hemodynamic metric for understanding the rupture risk of intracranial aneurysms. Several investigations have reported in the literature corresponding flow instabilities using established computational fluid dynamics tools. In this study, the occurrence of flow fluctuations is investigated using either Newtonian or non-Newtonian fluid models in patient-specific intracranial aneurysms using high-resolution lattice Boltzmann method simulations. Flow instabilities are quantified by computing power spectral density, proper orthogonal decomposition and spectral entropy, and fluctuating kinetic energy of velocity fluctuations. Furthermore, these hemodynamic parameters are compared between the ruptured and unruptured aneurysms. Our simulations reveal that the pulsatile inflow through the neck in a ruptured aneurysm is subject to a hydrodynamic instability leading to high-frequency fluctuations around the rupture position throughout the entire cardiac cycle. At other locations, the flow instability is only observed during the deceleration phase; typically, the fluctuations begin there just after peak systole, gradually decay, and the flow returns to its original, laminar pulsatile state during diastole. In the unruptured aneurysm, there is only minimal difference between Newtonian and non-Newtonian results. In the ruptured case, using the non-Newtonian model leads to a considerable increase of the fluctuations within the aneurysm sac.
\end{abstract}

\maketitle
\section{\label{sec:level1}Introduction}
\indent Intracranial aneurysms are local dilatations in cerebral arteries, highly prevalent in adults ($\sim$3.3\%)~\cite{brisman2006cerebral,vlak2011prevalence}. With the development of non-invasive medical imaging techniques, unruptured intracranial aneurysms are being increasingly detected. While some patients may remain healthy throughout their life, the risk of rupture stays present in connection with aneurysm growth. The rupture, in turn, can be fatal or often lead to severe disabilities~\cite{suarez2006aneurysmal,hop1997case}. Most treatments for aneurysms involve surgical interventions, which may result in complications and sometimes even increase again the risk of aneurysm rupture. This highlights the need for a better \emph{a priori} evaluation of rupture risk. To address this, a number of attempts have been made to systematically quantify rupture risk. E.g., Greving et al.~\cite{greving2014development} developed a risk score system (PHASES) used to predict the risk of aneurysm rupture based on various patients and aneurysm characteristics, such as smoking habits, size and location of aneurysms, etc. Despite lower risk scores indicating that there is no need for further treatment, such aneurysms may still rupture, pointing to the need for further research to better understand and predict the rupture risk and improve patient treatment outcomes.\\
\indent It has been argued that hemodynamics play an important role in the initiation, growth, and subsequent rupture of intracranial aneurysms~\cite{cebral2011association,Meng_2014_High_WSS_Low_WSS}. The presence of oscillatory shear stress distributions and complex flow patterns within the aneurysm sac has been hypothesized to be a potential contributor to the rupture of aneurysms~\cite{jain2016transitional,kataoka1999structural}. The existence of such unsteady flow structures is currently a hot topic of discussion within the community. Advanced in vivo experiments and numerical studies observing non-laminar blood flow behavior associated with intracranial aneurysms have been documented in the literature~\cite{ford2008piv,evju2013study,baek2009wall,steinman2013variability}. Kurokawa et al.~\cite{kurokawa1994noninvasive} reported sound at frequencies in the range of 150-800~Hz caused by flow-induced vibration of the aneurysm wall. Steiger et al.~\cite{steiger1986low} reported low-frequency flow fluctuations in cerebral saccular aneurysms of six patients out of twelve using Intraoperative Doppler recording. A spectral analysis and concomitant glass model studies showed a correlation between flow fluctuation and aneurysm type. Unexpected vibrations were also reported in experimental aneurysm studies, which are possibly associated with fluctuating flows ~\cite{ferguson1970turbulence,simkins1974vibrations,steiger1987basic,sekhar1984noninvasive,sekhar1990acoustic}. The significance of fluctuations in blood flow has been thoroughly explored and discussed by Steinman and his research team~\cite{antiga2009rethinking}. They found that significant narrow-band vibrations ranging from 100 to 500 Hz could be detected in two out of the three tested aneurysm geometries. This finding is in accordance with the study by Jain~\cite{jain2022effect}. Furthermore, the pulsatile nature of blood flow has been observed to influence aneurysm stability in both numerical and experimental studies~\cite{straatman2002hydrodynamic,blackburn2007instability}. It is indeed known that a pulsatile flow can become transitional in a straight pipe during the deceleration phase of the oscillation at Reynolds numbers below 2000 ~\cite{einav1993experimental}. This explains why several studies discovered indications of transitional flow in cerebral vessels, despite the low Reynolds numbers found there, and  suggesting an elevated risk of rupture~\cite{balasso2019high,valen2013high}. \\
\indent\Gls{cfd}, due to its non-invasive nature, is increasingly used to investigate the hemodynamics in patient-specific intracranial aneurysms. Numerous high-fidelity \gls{cfd} studies have been conducted to study, among other aspects, flow fluctuations in ruptured and unruptured intracranial aneurysms~\cite{cebral2011association,xiang2011hemodynamic,xu2018flow,khan2021prevalence,valen2014high}. From statistical analysis, they found that the occurrence of a rupture in aneurysms is correlated with abnormally high and low wall shear stresses, high oscillatory shear acting on the vessel wall, the presence of complex flow structures, and concentrated inflow jets. Baek et al.~\cite{baek2010flow} reported flow instabilities occurring during the deceleration phase of a cardiac cycle at the internal carotid artery and concluded that the occurrence of low-frequency fluctuations depended on the geometry. Ford et al.~\cite{ford2012exploring} and Valen-Sendstad et al.~\cite{valen2011direct,valen2013high} confirmed the presence of high-frequency fluctuations in patient-specific terminal and intracranial aneurysms. Still, many \gls{cfd} studies may be overlooking potentially relevant flow features such as high-frequency flow instabilities, simply due to inappropriate solver properties (highly dissipative schemes) and/or inadequate spatial and temporal resolutions, fluid models, or boundary conditions ~\cite{valen2014mind,khan2015narrowing}. 

As of today, a generalizable criterion for accurately assessing the risk of rupture in intracranial aneurysms remains elusive~\cite{chung2015cfd}. In an attempt to contribute to the discussion on the existence and properties of flow fluctuations, we consider in the present study two patient-specific intracranial aneurysms with known outcomes via direct numerical simulations (\gls{dns}, i.e., resolving properly all physical scales in time and space): one remaining unruptured, and the other one rupturing at a known location. Additionally, both Newtonian and non-Newtonian fluid models will be considered in the \gls{dns}.

This comparative study can be used to provide valuable information regarding the underlying mechanisms and possible consequences on hemodynamics. The article is organized as follows. \Cref{sec:numericalMethod} introduces the numerical model used to conduct the simulations and the involved analysis methods. \Cref{sec:level3Configuration} details the geometric models and parameters of the different cases, and \Cref{sec:results} presents the simulation results. We summarize our findings and discuss the limitations of this study in \Cref{sec:discussion}.

\section{Theoretical background\label{sec:numericalMethod}}
\subsection{Lattice Boltzmann method}
The \gls{lbm} is an alternative to classical approaches to solve the Navier-Stokes equation relying on the kinetic theory of gases. Instead of directly solving the classical mass and momentum balance equations, a system of coupled hyperbolic equations representing the balance of distribution functions and a truncated version of the Boltzmann equation is solved. This system of equations is readily shown to recover the Navier-Stokes equation in the hydrodynamic limit. The final form of the discrete space/time-evolution equation is:
 \begin{equation}
 f_i(\bm{r}+\bm{c}_i\delta_t, t+\delta_t) - f_i(\bm{r}, t) = \frac{\delta_t}{\bar{\tau}}\left(f_i^{\rm eq}(\rho,\bm{u}) - f_i(\bm{r}, t)\right),
 \label{eq:LBGKE}
 \end{equation}
where $f_i$ are discrete distribution functions, $\bm{c}_i$ corresponding discrete velocities, $\bar{\tau}$ is the relaxation time determining the fluid viscosity, $\delta_t$ the time-step size, and $f_i^{\rm eq}$ the equilibrium distribution function.

In this form of the solver, also known as the single relaxation time model, the Fourier number, i.e., non-dimensional viscosity, has a strong effect on the stability. It was reported that this method is basically unstable when the Fourier number is less than 0.005 or larger than 0.6~\cite{hosseini2019stability,lallemand2000theory}, which limits its applications. To overcome that issue many different and more advanced collision models have been developed over the years ~\cite{d2002multiple,geier2006cascaded,latt2006lattice}. 

In the present study, we will use a specific collision operator that was presented in earlier work and validated for medical flows with both Newtonian and non-Newtonian properties: A modified version of a central Hermite-based multiple relaxation time operator allowing for independent control of the dissipation rate of normal modes. This collision model has been shown to dramatically improve the stability domain of the lattice Boltzmann solver. Interested readers are referred to our published paper~\cite{hosseini2022lattice} for detailed discussions. In this model, the collision process is carried out in the space of Hermite central moments:
\begin{equation}
 f_i(\bm{r}+\bm{c}_i\delta_t, t+\delta_t) - f_i(\bm{r}, t) = \bm{\mathcal{T}}^{-1}\bm{S} \bm{\mathcal{T}} \left(\bm{f}^{\rm eq} - \bm{f}\right),
 \label{eq:MRT}
 \end{equation}
where here $\bm{\mathcal{T}}$ and $\bm{\mathcal{T}^{-1}}$ are the moments transform matrix and its inverse, respectively. For details on the transformation matrix readers are referred to Hosseini et al.~\cite{hosseini2021central,hosseini2020compressibility}. The diagonal matrix $\bm{S}$ indicates the relaxation rate of different moments. Here, for optimal stability, all relaxation rates except those tied to shear viscosity are set to one.
\subsection{Non-Newtonian model}
The viscosity of blood is constant in the Newtonian approximation while it is a function of shear rate, $\dot{\gamma}$, in non-Newtonian models. A wide variety of non-Newtonian models can be found in the literature such as the commonly used Carreau-Yasuda~\cite{bernabeu2013impact}, Casson~\cite{morales2013newtonian} and Cross~\cite{hosseini2022lattice} models. However, the viscosity approaches infinity when the shear rate is close to zero in the Casson model~\cite{boyd2007analysis}, which is obviously nonphysical. Fortunately, both Cross and Carreau-Yasuda models are capable of predicting the Newtonian plateaus at either very low or high shear rate regions. In this work, the Cross non-Newtonian model has been used to describe the blood rheology when non-Newtonian effects are taken into consideration~\cite{kehrwald2005lattice}. Note that it is straightforward to exchange the viscosity model in the \gls{lbm}.\\

The modified Cross model describing the shear-thinning behavior of blood is formulated as~\cite{cross1965rheology}:
\begin{equation}
    \mu(\dot{\gamma})= \mu_{\infty} + \frac{\mu_0 - \mu_{\infty}}{ {\left[1 + {(\lambda\dot{\gamma})}^n\right]} },
\end{equation}
where $\mu_0$  and $\mu_{\infty}$ are asymptotic values of viscosity at very low and high shear rates, respectively. The model involves also $\lambda$, a constant parameter with the dimension of time, and $n$ as a dimensionless constant that can govern the degree of shear thinning. In the present case,  $\mu_{\infty}$ = 0.0036~Pa.s, $\mu_0$ = 0.126~Pa.s, $\lambda$ = 8.2~s, $a$ = 1 (and therefore does not even appear in the equation) and $n$ = 0.64~\cite{gambaruto2013shear,bodnar2009numerical,campo2015review}.
The local shear rate $\rm{\dot{\gamma}}$ is defined as: 
\begin{equation}\label{eq:gammadot}
    \dot{\gamma} = \sqrt{2D_{II}},
\end{equation}
where $D_{II}$ is the second invariant of the deviatoric rate-of-strain tensor. The invariants of a rank-two tensor $A$ are defined as the coefficients of the corresponding characteristics polynomial, i.e.
\begin{equation}
    {\rm det}(A - \lambda I) = - \lambda^3 + D_{I} \lambda^2 + D_{II} \lambda + D_{III},
\end{equation}
where $D_{I}$, $D_{II}$ and $D_{III}$ are the first, second and third invariants of $A$. The first invariant is readily found as the trace of $A$, while the second invariant is:
\begin{equation}
    D_{II} = \frac{1}{2}\left[ {\rm tr}(A^2) - {\rm tr}(A)^2 \right].
\end{equation}
In the case of a trace-free tensor $A$ the first invariant would reduce to zero while the second one would yield:
\begin{equation}
    D_{II} = A:A,
\end{equation}
where $A:A$ designates the component-wise Frobenius inner product, i.e., the sum of the Hadamard product of the two tensors. This last equality is obtained by virtue of:
\begin{equation}
    {\rm tr}(A A^\dagger) = A:A^\dagger,
\end{equation}
and noting that the stress tensor is symmetrical, i.e.,
\begin{equation}
    A = A^\dagger.
\end{equation}
Now denoting the stress tensor as $S$ the second invariant is readily obtained as:
\begin{equation}\label{eq:DII-2D}
    D_{II} = \frac{1}{2}\left(S-\frac{1}{D}{\rm trace}(S)\right):\left(S-\frac{1}{D}{\rm trace}(S)\right),
\end{equation} 
where $D$ is the physical dimension of the considered system.

The rate-of-strain, in the context of the \gls{lbm}, is computed locally from discrete distribution functions~\cite{hosseini2022lattice}:
\begin{equation}\label{eq:rate_LB}
    S = -\frac{1}{2\rho\bar{\tau}c_s^2}\sum_i \bm{c}_i\otimes\bm{c}_i \left(f_i - f_i^{\rm eq}\right),
\end{equation}
where $\rho$ is the density.  
\subsection{Fluctuating kinetic energy}
In the context of the present study,  we will focus the majority of our analysis on fluctuating components of the velocity field. To quantify the variations, the instantaneous velocity is decomposed into a mean component $\overline{\bm{u}}(\bm{r},t)$ and a fluctuating component $\bm{u}'(\bm{r},t)$ where:
\begin{equation}
    \overline{\bm{u}}(\bm{r},t) = \int_{t_0}^{t_0+\Delta t} \bm{u}(\bm{r},t) dt,
\end{equation}
where $\Delta t$ is the averaging period and:
\begin{equation}
    \bm{u}'(\bm{r},t) = \bm{u}(\bm{r},t) - \overline{\bm{u}}(\bm{r},t).
\end{equation}
\gls{fke}, a variable we will be using quite frequently, is expressed as:
\begin{equation}
    \mathrm{FKE}(\bm{r},t) = \frac{1}{2}\int_{t_0}^{t_0+\Delta t} \bm{u}'(\bm{r},t)\bm{u}'(\bm{r},t) dt.
\end{equation}
\subsection{Power spectral density}
\Gls{psd} shows how the power of a signal is distributed over the frequency domain; this provides a quantitative measure of flow fluctuations.  
Here, the \gls{psd} at each probe point has been computed using Matlab (Welch's method).

The discrete-time power spectral density ($S_{xx}(f)$) is defined at frequency $f$ by~\cite{ricker2003echo}:
\begin{equation}
    S_{xx}(f) = \frac{\delta_t^2}{\Delta t}\left|\sum_{t=t_0}^{t_0+\Delta t}\phi(t) \exp{\left[-2\pi\sqrt{-1}f (t-t_0)\right]} \right|^2
\end{equation}
where $\delta t$ and $\Delta t$ are the sample interval and  the length of the signal, respectively, while $\phi(t)$ is the velocity signal.
\subsection{Spectral entropy}
In order to quantify the temporal flow instability in the velocity field of interest, the spectral entropy $S_d$ is computed by a Proper Orthogonal Decomposition (POD) of the velocity~\cite{abdelsamie2017spectral}. The flow regimes can then be distinguished into stable or unstable flows based on spectral entropy. The relative energy in the aneurysm $P_i$ is obtained from the corresponding eigenvalue $\lambda_i$, which quantifies the energy content of the $i_{th}$ mode:
\begin{equation}
    P_i = \frac{\lambda_i}{\sum_{j=1}^N \lambda_j}
\end{equation}
where $N$ is the total number of modes in the analysis.

The spectral entropy measures the average energy distributed in the aneurysm across the $N$ modes, which can be used to quantify the stability of hemodynamic flows. The definition is as below:
\begin{equation}
    S_d = -\sum_{i=1}^N P_i \ln{P_i}
    \label{pod_eq}
\end{equation}
Based on Eq.~(\ref{pod_eq}), the spectral entropy reaches its maximal value when the energy is evenly distributed over all the $N$ modes. Entropy is minimized when the velocity signal contains only one single mode, meaning a steady flow ($S_d=0$).

\section{Configurations and numerical set-up\label{sec:level3Configuration}}
\subsection{Case studies}
\begin{figure*}[!hbt]
	\centering
	\includegraphics[width=1.5\columnwidth]{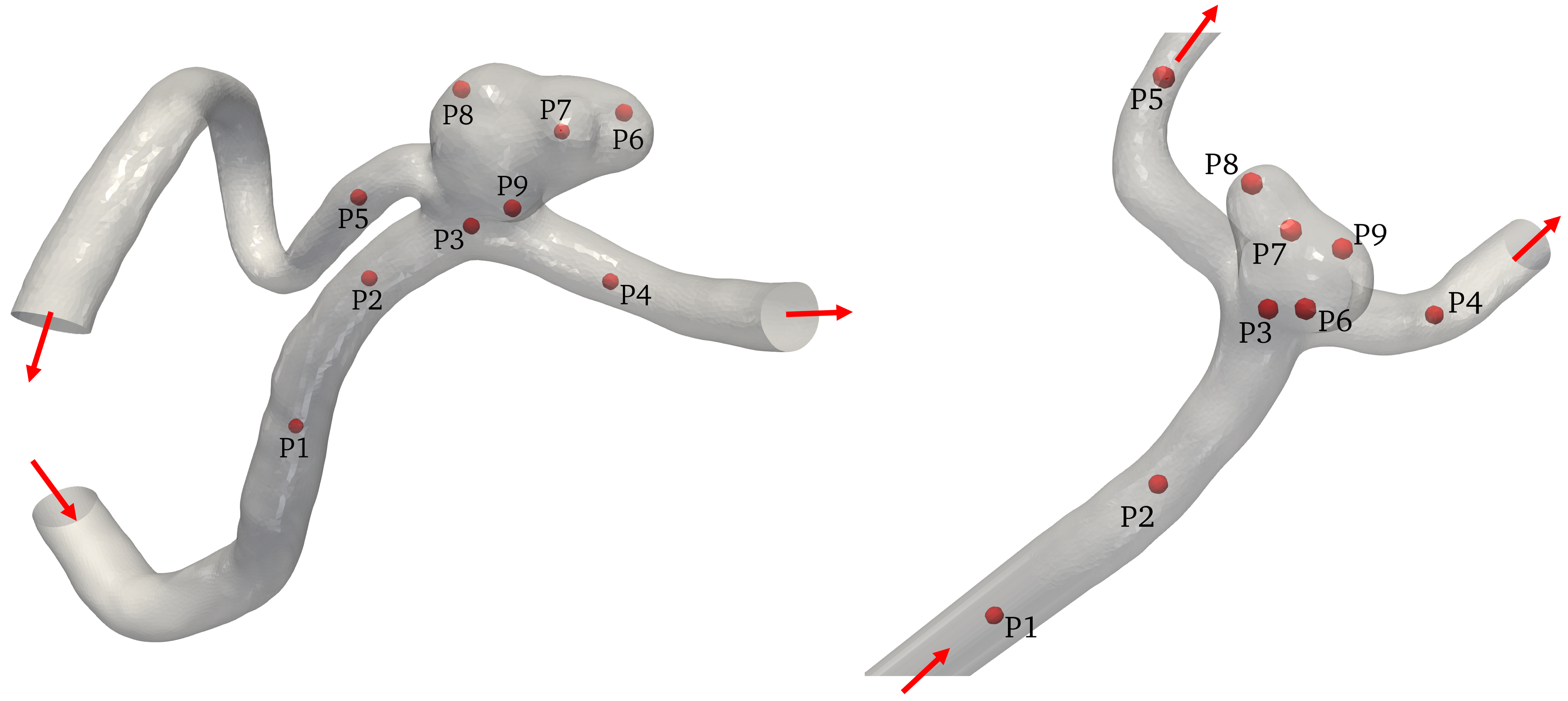}
\caption{Representation of the patient-specific aneurysm models employed in simulations (Left: ruptured case, Right: unruptured case). The red arrows point to the single inlets and double outlets, respectively. Note that point P6 corresponds to the ruptured location on the left aneurysm.}
\label{Fig:rup_Unrup_geoemtry}
\end{figure*}

In the present study, hemodynamic features are investigated in two patient-specific intracranial aneurysm geometries, shown in Fig.~\ref{Fig:rup_Unrup_geoemtry}. The specificity of these two cases is that one is known not to have ruptured, while the other has ruptured at a known location within the aneurysm sac. 
The unruptured aneurysm (of size 1.6~mm) was found in the middle cerebral artery (MCA) in a 53-year-old female patient who experienced an acute subarachnoid hemorrhage and severe headaches  and was treated by coiling. The three-dimensional rotational angiography (3DRA) was performed on an Artis Q angiography system (Siemens Healthineers AG, Forchheim, Germany) with a spatial resolution of 0.28 $\times$ 0.28 $\times$ 0.28~\rm{mm} to produce tomography slices of the volume of interest. Subsequently, the reconstructed raw image was obtained from a syngo X Workplace (Siemens Healthcare GmbH, Forchheim, Germany) through an 'HU auto' kernel. For more details, we refer readers to~\cite{voss2019multiple}. The ruptured aneurysm was detected in the posterior inferior cerebellar artery (PICA) in a 51-year-old female patient. The 3D-DSA scan was carried out on an AXIOM-Artis (Siemens Healthineers AG, Forchheim, Germany) with the same spatial resolution (0.28~mm) as in the other case.  The raw images were reconstructed using 'EE/auto' kernel (Siemens Healthcare GmbH, Forschheim, Germany).\\
\indent The geometric models of the aneurysms were reconstructed on this basis and further processed to simplify numerical simulations, retaining only the intracranial aneurysms and the corresponding major arteries to guarantee reasonable inflow and outflow, while the distal arterial segments were cut off.
\subsection{Numerical simulations\label{sec:benchmark}}
\begin{figure}[!ht]
	\centering
\includegraphics[width=0.65\columnwidth]{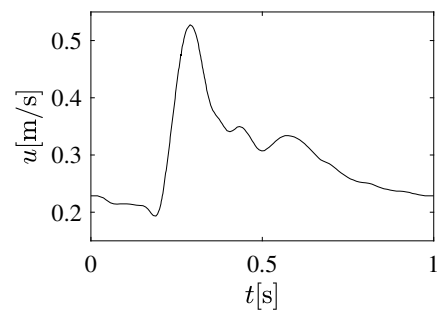}
\caption{Illustration of the inflow velocity waveform over an entire cardiac cycle imposed at the inlet.}
\label{Fig:InletVelocity_Profile}
\end{figure}

Simulations have been carried out using our in-house \gls{lbm} solver ALBORZ, which has already been thoroughly validated using a variety of benchmark studies~\cite{huang2022simulation,hosseini2021central,hosseini2019hybrid}. Blood was assumed to be an incompressible fluid with a density of 1000~kg/$\hbox{m}^3$ and, for Newtonian cases, kinematic viscosity equal to 4 $\times$ $10^{-6}~\hbox{m}^2$/s. For the non-Newtonian case, the modified Cross model was used to describe the blood rheological properties. For each configuration considered in the present work, simulations have been carried out using three different resolutions, allowing to make sure extracted results are converged physical data, as opposed to numerical artifacts such as Gibbs oscillations. To make sure simulations are converging to the incompressible limit and keep computation costs as low as possible, the time-step size refinement follows an acoustic scaling resulting in a fixed \gls{cfl} number, set to a low value of $0.052$. The resulting time-step, grid size, and resulting \gls{cfl} number are listed in Table~\ref{table:summaryNonNewtoianIdeal}.
\begin{table}[th]
\centering
\footnotesize
\setlength{\tabcolsep}{3pt}
\caption{Summary of numerical simulation parameters used in this study.}
\begin{tabular}{||l|l|l|l||}
\hline
 Resolution & $\delta_r$[m] & $\delta_t$[s] & \gls{cfl}\\
\hline\hline
\gls{lr} & 2.6$\times10^{-4}$  &  7.6$\times10^{-6}$  & 0.052\\
\hline
\gls{mr} & 1.3$\times10^{-4}$  & 3.8$\times10^{-6}$ & 0.052\\
\hline
\gls{hr} & 6.5$\times10^{-5}$  & 1.9$\times10^{-6}$ & 0.052\\
\hline
\hline
\end{tabular}
\vspace{-4mm}\label{table:summaryNonNewtoianIdeal}
\end{table}
The period of one cardiac cycle is set to 1~$\rm{s}$ for both considered cases, and the corresponding inflow velocity waveform is presented in Fig. \ref{Fig:InletVelocity_Profile}. The inlet boundary condition was set as a pulsatile velocity profile, while the pressure at the outlets was kept constant. The walls were assumed to be rigid with no-slip boundary conditions. All simulations were run for 3 cycles, and data sampling was performed during the third cycle to remove the influence of initial transients.
\begin{figure*}[!ht]
	\centering
    \begin{subfigure}[Ruptured case]{\includegraphics[width=0.49\linewidth]{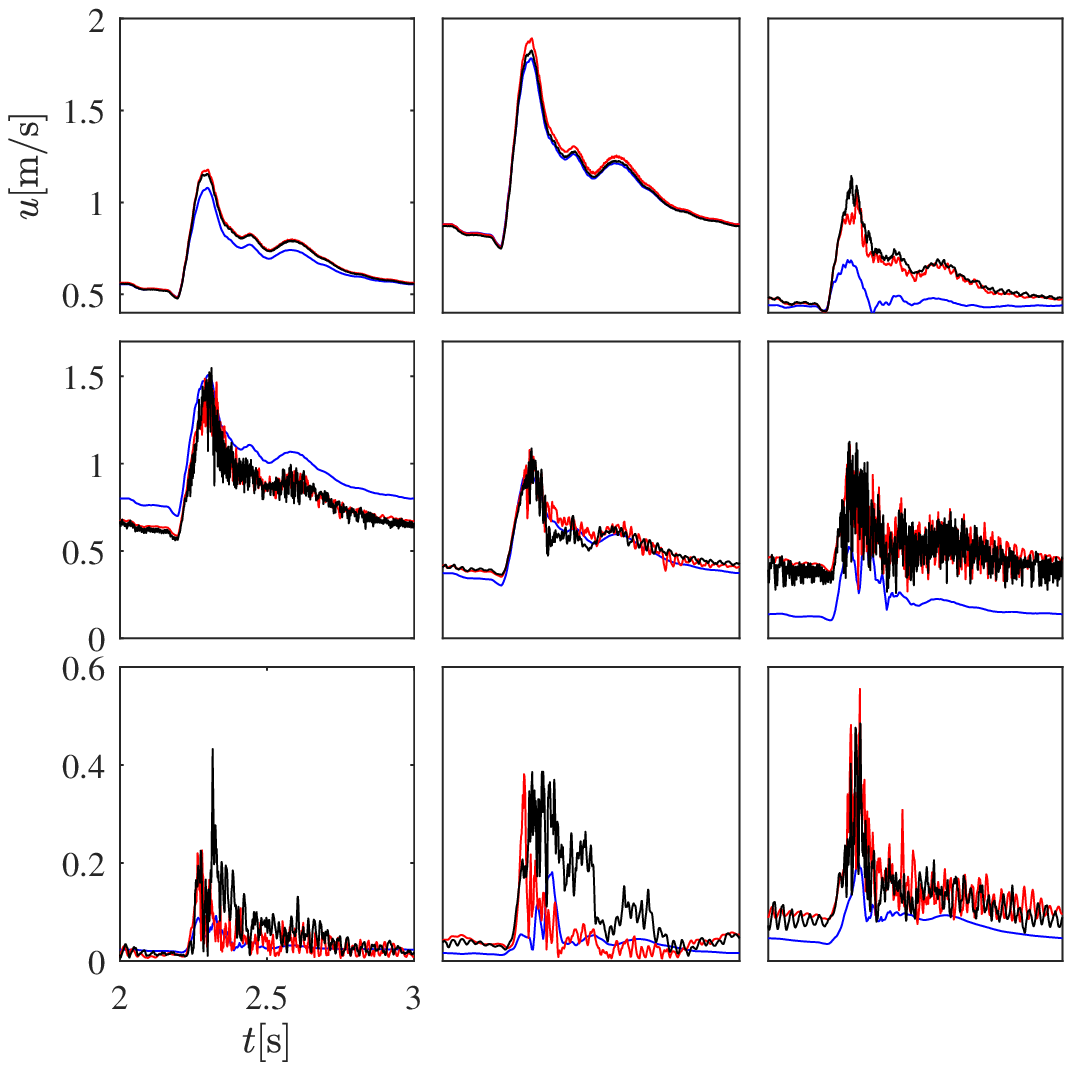}}
        \centering 
        \label{fig:ruptured_grid}
    \end{subfigure}
    \begin{subfigure}[Unruptured case]{\includegraphics[width=0.49\linewidth]{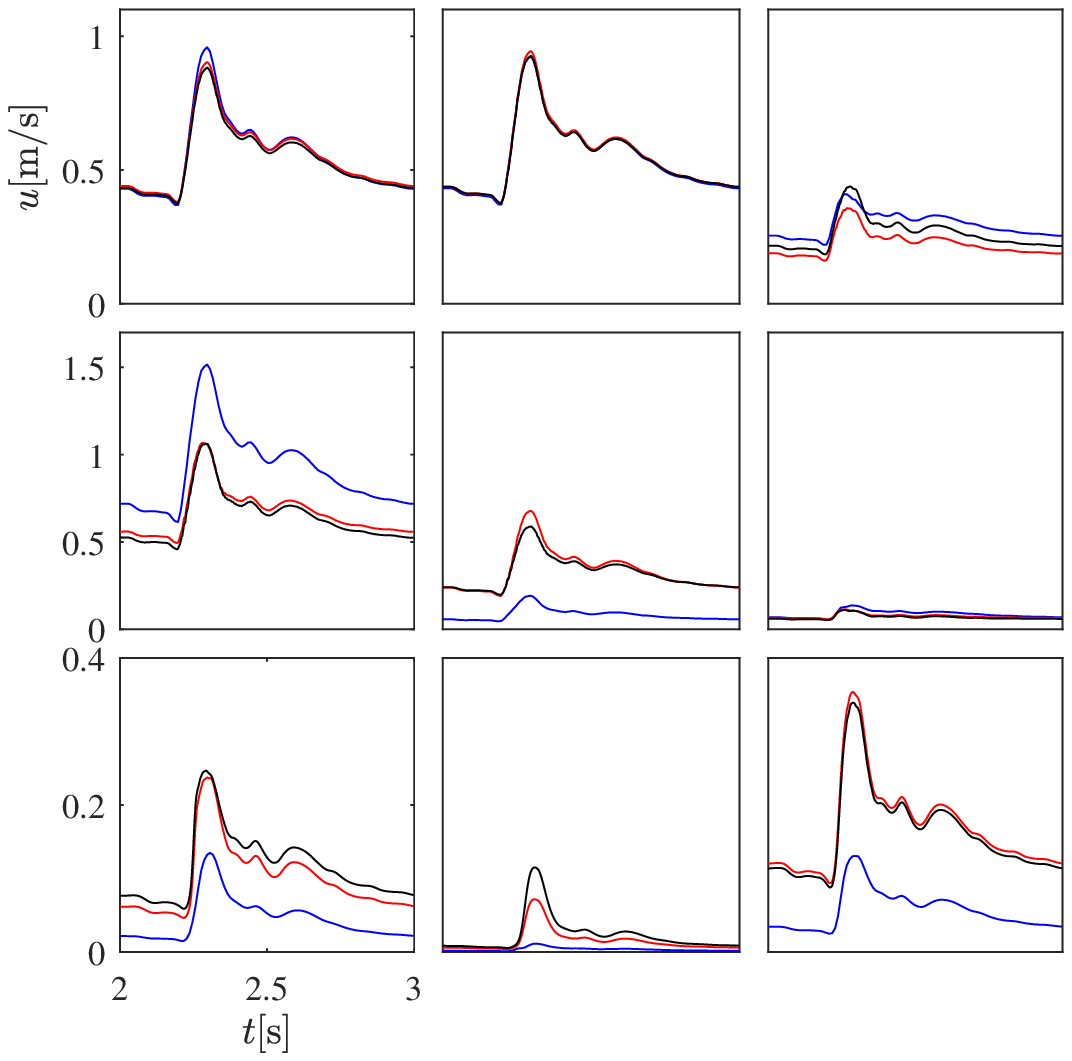}}
        \centering 
        \label{fig:unruptured_grid}
    \end{subfigure}
\caption{Time histories of the simulated velocity in both aneurysms at 9 monitoring points during the last cardiac cycle. The blue, red and black solid lines denote the flow velocity for increasing resolution (\gls{lr}: low, \gls{mr}: middle, and \gls{hr}: high, respectively). From top left to bottom right, results at points P1 to P9 are shown.}
\label{Fig:RuptureNewtonian_Fluctuation}
\end{figure*}
The flow values at up to 9 points were tracked upstream and downstream of the aneurysm, as well as within the aneurysm sac, in order to monitor any manifestation of possible flow fluctuations. It is worth noting that the inlets and outlets of both models were extended along the normal vector of the corresponding surfaces by 5 inlet diameters to reduce possible effects of the inlet and outlet boundary conditions. 
All results, with a focus on fluctuations, are now detailed in the next section.
\section{Results\label{sec:results}}
\subsection{Impact of resolution}
In order to obtain accurate results while keeping acceptable computational costs, we first conducted a grid-independence study on the ruptured and unruptured aneurysms considering Newtonian simulations. The grid size and time-step were varied simultaneously (keeping \gls{cfl} number constant) from 260~$\mathrm{\mu m}$ down to 65~$\mathrm{\mu m}$ and from 7.6~$\rm{\mu s}$ to 1.9~$\rm{\mu s}$, respectively.

\begin{figure*}[!ht]
	\centering
    \begin{subfigure}[Ruptured case]{\includegraphics[width=0.49\linewidth]{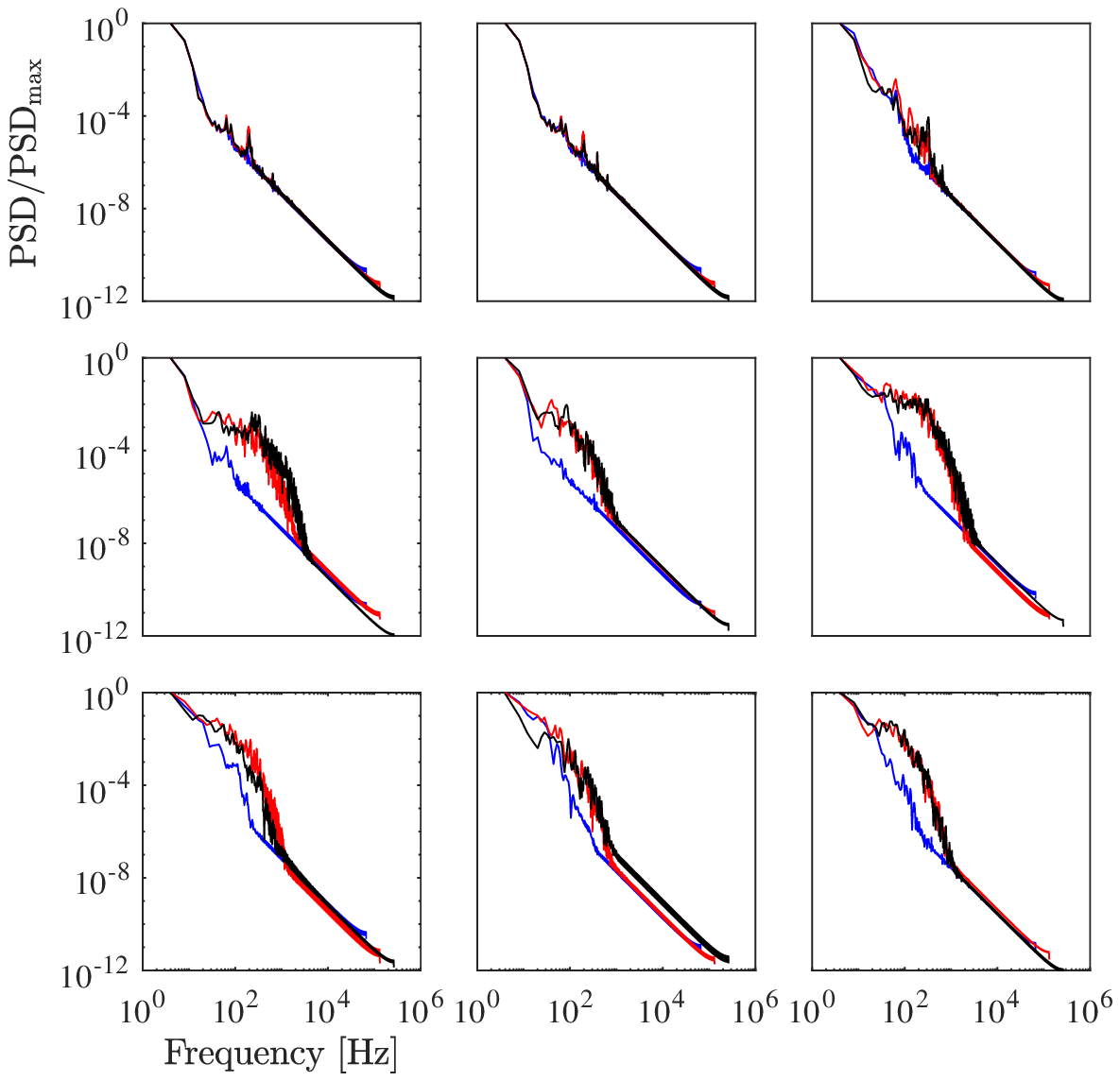}}
    \end{subfigure}
    \begin{subfigure}[Unruptured case]{\includegraphics[width=0.49\linewidth]{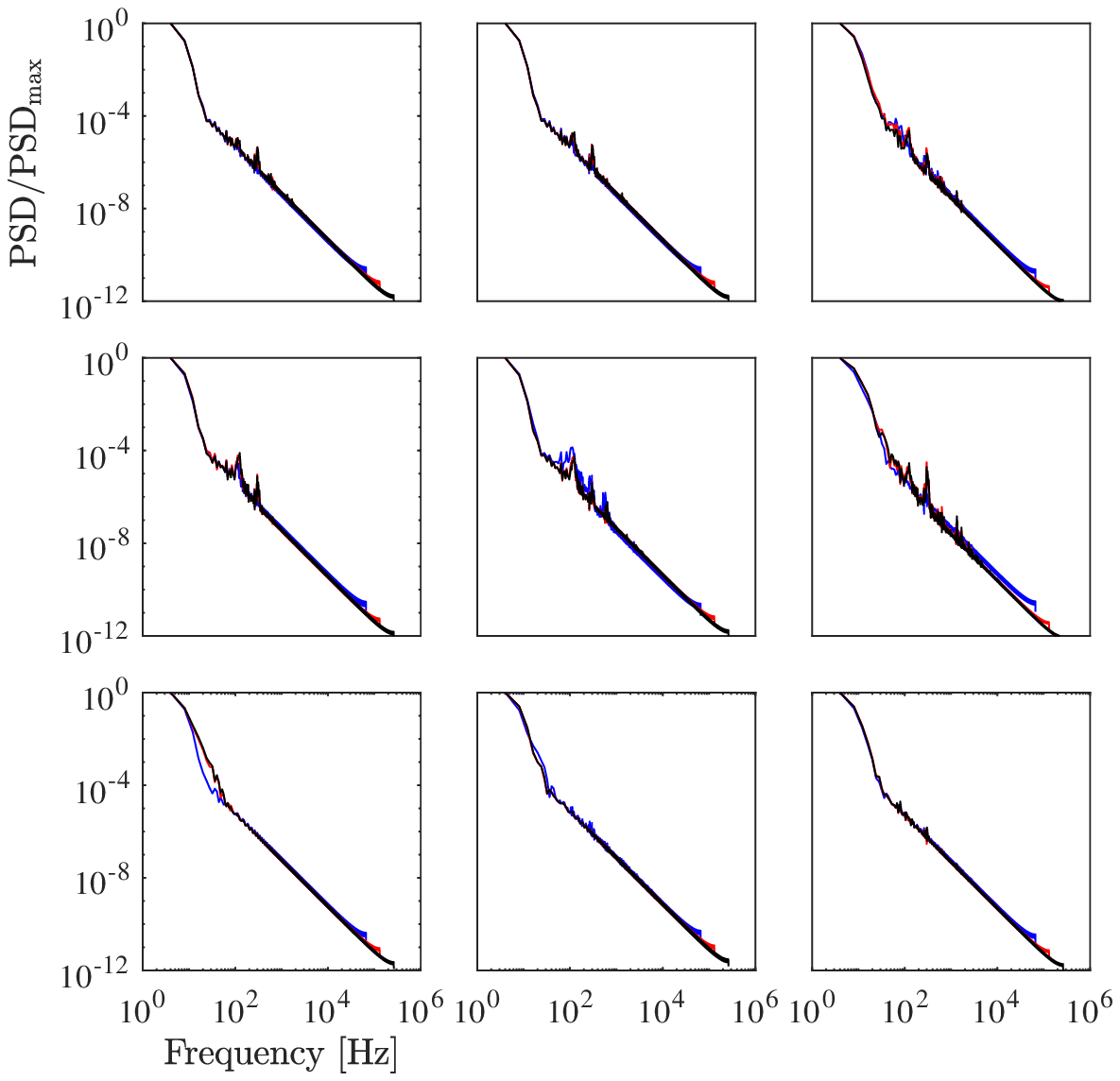}}  
    \end{subfigure}	
\caption{Energy spectra of the velocity distributions from Fig.~\ref{Fig:RuptureNewtonian_Fluctuation} for both aneurysms. The blue, red and black solid lines denote again the flow velocity for increasing resolution (\gls{lr}: low, \gls{mr}: middle, and \gls{hr}: high, respectively). From top left to bottom right, results at points P1 to P9 are shown.}
\label{Fig:PSD_resolution_Rup_Unrup}
\end{figure*}

As expected, the high-frequency fluctuations were found to be very sensitive to the resolution in time and space. 
As shown in Fig.~\ref{Fig:RuptureNewtonian_Fluctuation}, the low-resolution simulation (\gls{lr}) failed to predict any flow fluctuation in the entire domain for both cases. It also showed pronounced discrepancies in both aneurysms when compared to results at higher resolutions. Therefore, the choice of a suitable resolution plays a significant role in obtaining proper data regarding high-frequency oscillations. On the other hand, both intermediate (\gls{mr}) and high-resolution (\gls{hr}) simulations exhibit high-frequency oscillations for exactly the same conditions and locations. Such fluctuations are found nowhere in the unruptured case in Fig.~\ref{fig:unruptured_grid}, while they are clearly visible at points P3 to P9 in the ruptured case. Additionally, the spectra obtained with both resolutions \gls{mr} and \gls{hr} show very similar features, as revealed in Fig.~\ref{Fig:PSD_resolution_Rup_Unrup}. Conversely, the \gls{lr} spectra reveal a lower energy content for all cases. The energy level is systematically lower in the unruptured aneurysm compared to the ruptured case. 

Still, minor discrepancies are still visible between \gls{mr} and \gls{hr} for the ruptured case at points P7 and P8, within the aneurysm sac. Those might be partly explained by the fact that the monitoring point was always pinned to the closest grid point, therefore, changing very slightly the actual position from one resolution to the other. It might also indicate that resolution \gls{hr} is still not fully sufficient to get grid-independence. However, it must be kept in mind that the central purpose of this study is to detect high-frequency fluctuations in cases where they appear. In that regard, the results obtained with \gls{mr} and \gls{hr} are fully identical, revealing such fluctuations at the same positions for the same conditions and with similar spectra. Considering that the total computational load is high but still acceptable using High-Performance Computers for the highest resolution (\gls{hr}), it has been finally kept for the rest of this paper in order to maximize accuracy.
\subsection{Comparison of flow features in the Newtonian case}
\subsubsection{Flow patterns in the sac}
\begin{figure*}[!ht]
	\centering
	\includegraphics[width=2\columnwidth]{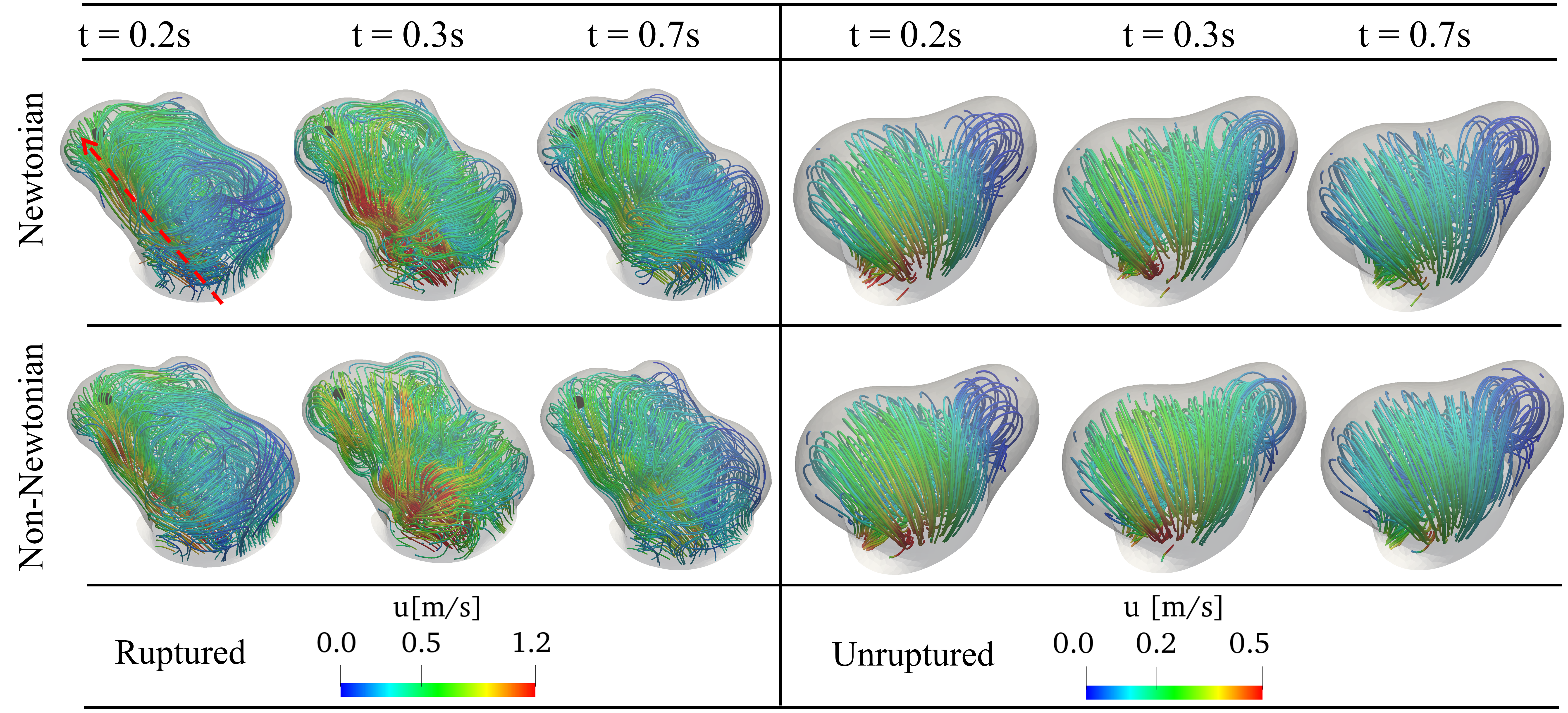}
\caption{Flow streamlines colored by flow velocity inside the aneurysm sac for the ruptured case (left column) and the unruptured case (right column) at the acceleration, peak, and deceleration systole, respectively. The black point and red arrow indicate the rupture location and high-speed jet, respectively. The top row of the figure shows results for the Newtonian model, bottom row corresponds to the non-Newtonian (Cross) model.}
\label{Fig:RuptureUnrup_FlowField}
\end{figure*}

A qualitative comparison of the flow streamlines in ruptured and unruptured aneurysms is shown in Fig.~\ref{Fig:RuptureUnrup_FlowField}, illustrating the flow streamlines in the sac at acceleration, peak, and deceleration systole during the heart cycle in both Newtonian (top row) and non-Newtonian cases (bottom row).\\
\indent This first comparison reveals differences between flow streamlines in ruptured and unruptured cases. In the ruptured aneurysm, it can be observed that a high-speed, concentrated jet originating from the parent vessel at a specific angle impinges onto the opposing wall, leading to the formation of complex vortical structures and disturbed flow patterns.
The peak velocity exceeds 1.2~$\rm{m/s}$ at time 0.3~$\rm{s}$ during the cardiac cycle. In contrast, the unruptured case presents a quite stable flow pattern with a single, large-scale vortex observed without major modifications in all three cycles and at the three selected time-points during the computation. The maximum velocity is less than 0.5~$\rm{m/s}$ at peak systole, indicating a much smoother and laminar flow.\\
\indent One further interesting observation that can be made here is that while Newtonian and non-Newtonian results are visually quite similar for the unruptured case, they exhibit pronounced differences for the ruptured case, also concerning large-scale flow structures. Corresponding effects will be discussed in more detail in a later section.\\
\indent This first, qualitative analysis revealed a stronger flow with larger variations in time in the ruptured case. Therefore, velocity fluctuations in time appear to play an important role and will be quantified in what follows.
\subsubsection{Velocity time traces}
First, it is interesting to determine more precisely the occurrence and location of flow fluctuations in the aneurysms. Previous Fig.~\ref{fig:ruptured_grid} showed the time traces of velocity, in particular at probe P6 in the ruptured aneurysm (remember that point P6 is the -- known -- location of rupture in this case).

\begin{figure*}[!ht]
	\centering
  \includegraphics[width=1.6\columnwidth]{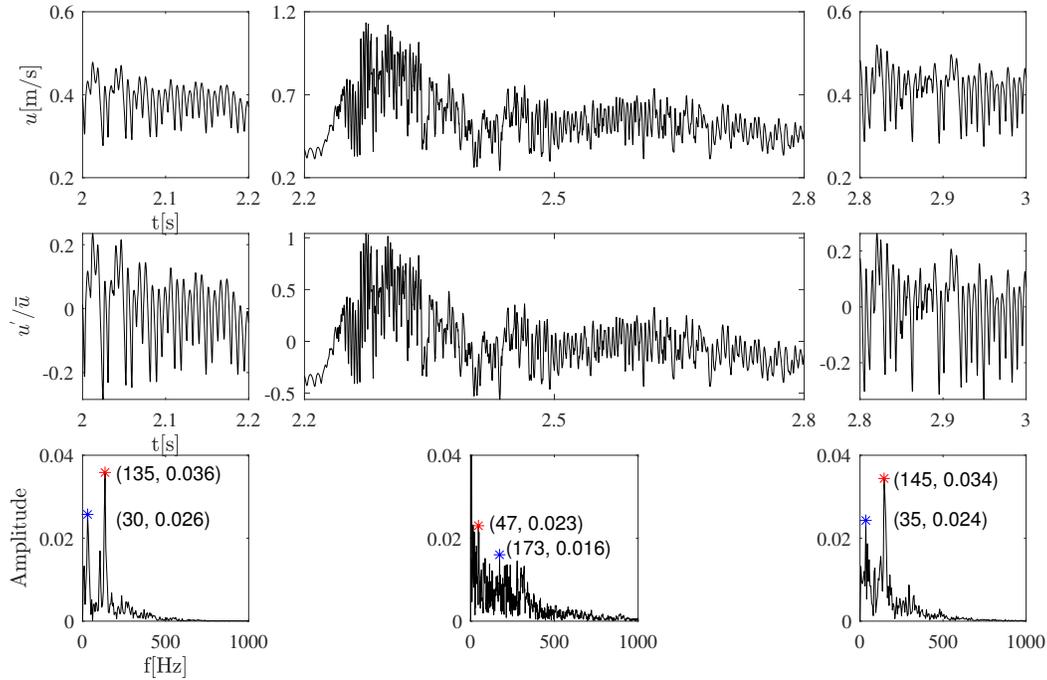}
\caption{Time history of velocity, normalized fluctuating velocity, and \gls{fft} of velocity at point P6 in the ruptured case during the last cardiac cycle (from 2 to 3~s), considering blood as a Newtonian fluid. Note that the figures in the two top rows use different vertical scales for a better readability. The red and blue stars in the bottom row indicate the first and second dominant frequencies.}
\label{Fig:RuptureNewtonianSeperated}
\end{figure*}

Looking back at Fig.\ref{fig:ruptured_grid}, it is noted that there are no obvious flow instabilities in the parent artery (points P1 and P2). Instead, flow fluctuations appear first at the bifurcation location (point P3), and are then visible in the entire sac and up to the two outlets (points P4$\&$P5). At bifurcation point P3, only weak fluctuations are observed, and only around peak systole. Le et al.~\cite{le2021dynamic} have also presented evidence of a highly dynamic inflow jet in the ostium of intracranial aneurysms. Through dynamic mode decomposition analysis, they have discovered that the high-frequency modes are of short duration and correspond to the flow separation at the proximal neck and the impingement of the jet onto the aneurysm wall. Within the aneurysm sac (points P6 to P9), both the frequency and amplitude of flow fluctuations are significantly higher (see Figs.~\ref{Fig:RuptureNewtonian_Fluctuation} and \ref{Fig:PSD_resolution_Rup_Unrup}). On the other hand, as the flow passes beyond peak systole, both the amplitude and frequency gradually decay, even if velocity instabilities persist over the entire cardiac cycle in part of the sac. Interestingly, the fluctuations are much stronger during the deceleration phase, which can be explained by the fact that transition effects are inhibited during the acceleration phase, and supported during the deceleration phase, as also observed in pulsating flows in straight tubes~\cite{juarez2003direct}. \\
\indent To better study the velocity fluctuations at the rupture point (P6), Fig.~\ref{Fig:RuptureNewtonianSeperated} was partitioned into three distinct phases within the last cardiac cycle (2-2.2~s, 2.2-2.8~s, 2.8-3.0~s), which allows for a better readability.
As can be seen from the first row of the plot, at the start of the cardiac cycle (left figure), the fluctuations are nearly uniform and the average velocity is around 0.4~m/s. Then, strong fluctuations appear after peak systole, at which the maximal velocity reaches almost 1.2~m/s. 
In the last stage, the flow instability decays toward the end of the deceleration phase, and the flow turns back into low-frequency fluctuations around an average velocity of 0.4~m/s (right figure), very similar to the results at the start of the cardiac cycle (compare left and right figures).

The second row exhibits the profile of normalized fluctuating velocity over time. It shows that the relative amplitude of the velocity fluctuations is around 20\% in the early and late phase of the cardiac cycle, while it goes up to 100\% around peak systole.

The lowest row presents the \gls{fft} of the fluctuating velocity signal. It can be observed that the dominating frequencies (red stars) are around 140~Hz in the early and late part of the cardiac cycle. During peak systole, the situation is far more complex because many frequencies are involved and the \gls{fft} signal is strongly populated, revealing the onset of turbulent conditions; the dominating frequency appears around 47~Hz, with a second one at 173~Hz. 
The first, low-frequency fluctuation can potentially be attributed to cycle-to-cycle fluctuations occurring when the flow undergoes deceleration, as previously reported in ~\cite{valen2011direct,xu2016exploring}.

Compared to the ruptured aneurysm, the unruptured model (not shown in the interest of space) exhibits no velocity fluctuations, with a very stable flow in the entire domain at all times. 

Based on the above results, it is concluded that high-frequency fluctuations are only detectable in the ruptured aneurysm. This observation is consistent, qualitatively, with an early computational study on intracranial aneurysms by Varble et al.~\cite{varble2016flow} and more recent works on that topic, e.g. ~\cite{behme2021069,gaidzik2021luminal}.
\subsection{Fluctuating kinetic energy}
\begin{figure}[!ht]
	\centering
	\includegraphics[width=0.8\columnwidth]{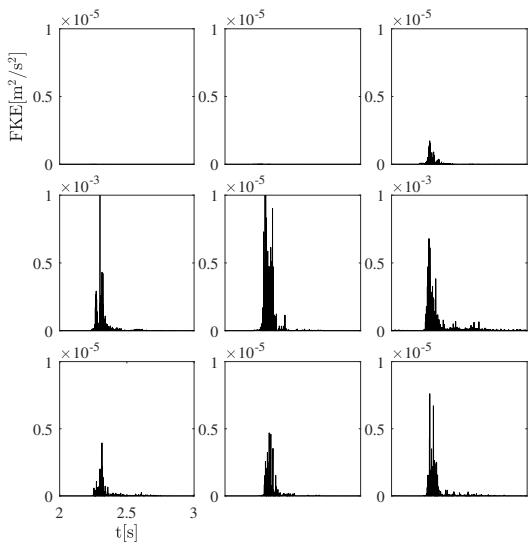}
\caption{Time histories of \gls{fke} in the ruptured case at all monitoring points during the last cardiac cycle, considering blood as a Newtonian fluid. From top left to bottom right, results at points P1 to P9 are shown. Note the different vertical scales for points P4 and P6.}
\label{Fig:FKE_RuptureNewtonian}
\end{figure}

\gls{fke} is an important tool to quantify flow fluctuations. Figure~\ref{Fig:FKE_RuptureNewtonian} presents the distribution of \gls{fke} at the monitoring points in the ruptured aneurysm case.

It can be seen that the time evolution of fluctuating kinetic energy is zero at any given time for points P1 and P2 in the parent artery, corresponding to the smooth velocity profiles and the absence of any fluctuations there. A low level of fluctuating kinetic energy ($\sim 2\times10^{-6}~\rm{m^2/s^2}$) is first observed at bifurcation point P3 around peak systole, revealing the onset of flow instability. A much higher level of fluctuating kinetic energy (orders of magnitude larger, $\sim 1\times10^{-3}~\rm{m^2/s^2}$) observed at the rupture point (P6) and within one outlet artery (P4) confirm the observations discussed previously. The substantial difference in fluctuating kinetic energy between the two outlets may be attributed to the particular shape of the ruptured aneurysm and vessel tree. In fact, the flow rate leaving through the outlet containing P4 is considerably larger than for the outlet containing P5. The maximum fluctuating kinetic energy is predominantly observed around peak systole. However, only at the rupture site (P6), noticeable fluctuations are also found during late systole, and the second-highest peak still reach $6\times 10^{-5}~\rm{m^2/s^2}$ there, which is higher than the overall peak value at all other points before and within the aneurysm sac.

The other points in the aneurysm sac (P7, P8, P9) show much lower \gls{fke} values, and those are only observed in the vicinity of peak systole; at the start and end of the cardiac cycle, \gls{fke} is basically 0 there. Overall, these very distinct features of \gls{fke} suggest that a strong flow instability as measured by fluctuating turbulent energy might be a characteristic feature of the rupture point within the aneurysm sac.
\subsection{Power spectral density}
\begin{figure}[!ht]
	\centering
	\includegraphics[width=0.9\columnwidth]{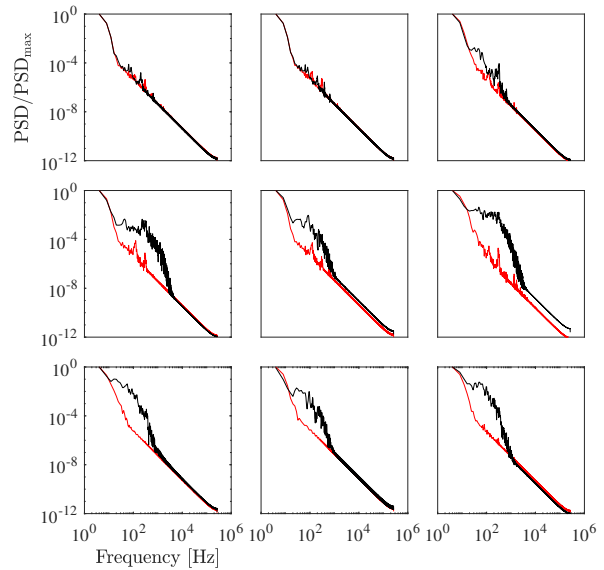}
\caption{Energy spectra of velocity in the ruptured (black) and unruptured aneurysm case (red) at the 9 monitoring points during the last cardiac cycle, considering blood as a Newtonian fluid. From top left to bottom right, results at points P1 to P9 are shown. }
\label{Fig:Rup_UnrupNewtonianPSD}
\end{figure}

Figure~\ref{Fig:Rup_UnrupNewtonianPSD} shows the energy spectra of the flow velocity at the monitoring points for both ruptured and unruptured case.
For all points within the aneurysm sac, the \gls{psd} of the ruptured case (black) are much larger compared to the unruptured condition (red).
In the unruptured case, the \gls{psd} of the flow velocity reveals a nearly monotonic decay as a function of frequency; no large peaks are observed. In contrast, the ruptured case shows distinct spectral characteristics corresponding to what is expected in a turbulent-like flow with strong fluctuations, with a large and broad increase of the \gls{psd} around approximately 100~Hz. The same observation applies for all points P4 to P9.

\begin{figure*}[!ht]
	\centering
    \begin{subfigure}[Ruptured case]{\includegraphics[width=0.4\linewidth]{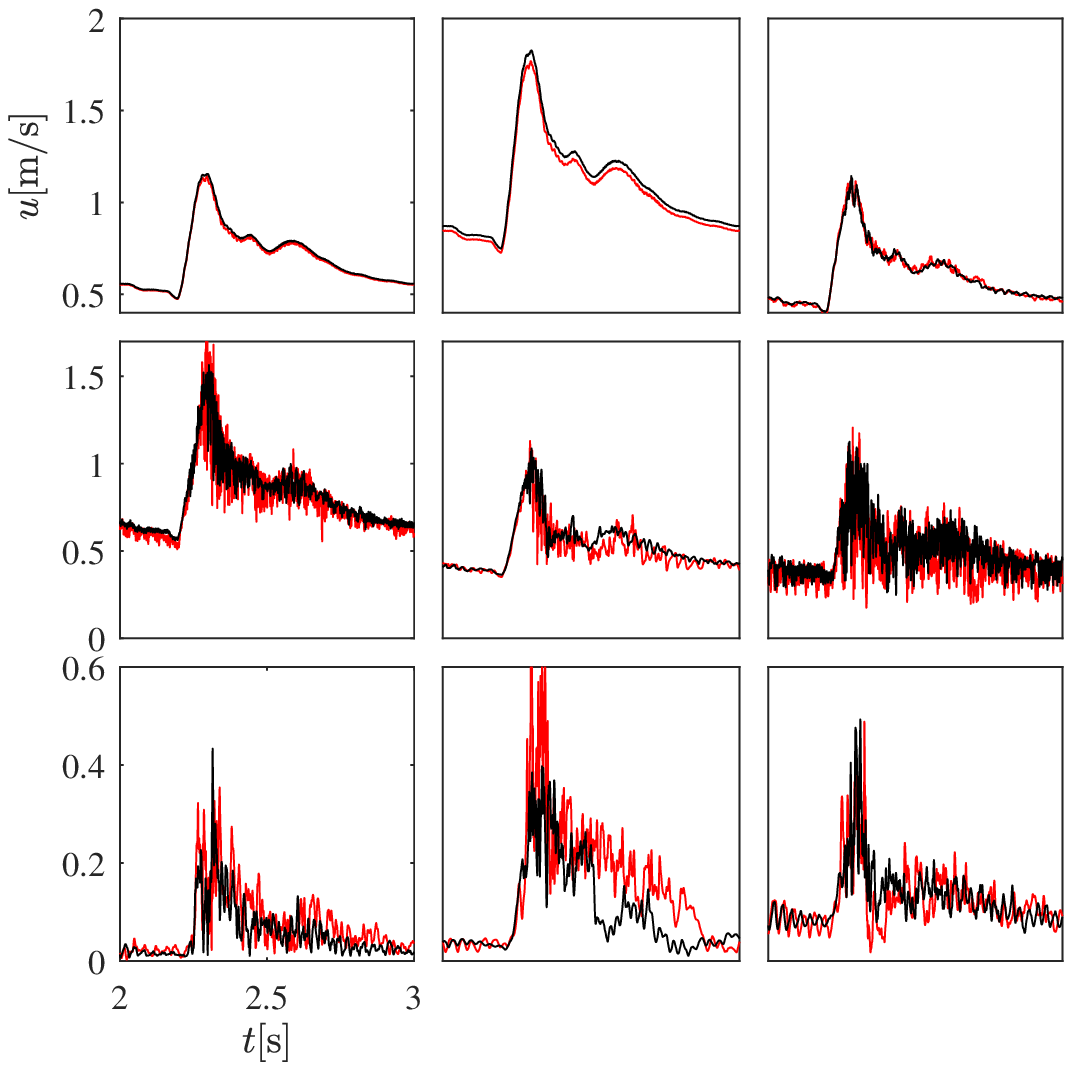}}
        \centering
    \end{subfigure}    
    \hspace{1cm}
    \begin{subfigure}[Unruptured case]{\includegraphics[width=0.4\linewidth]{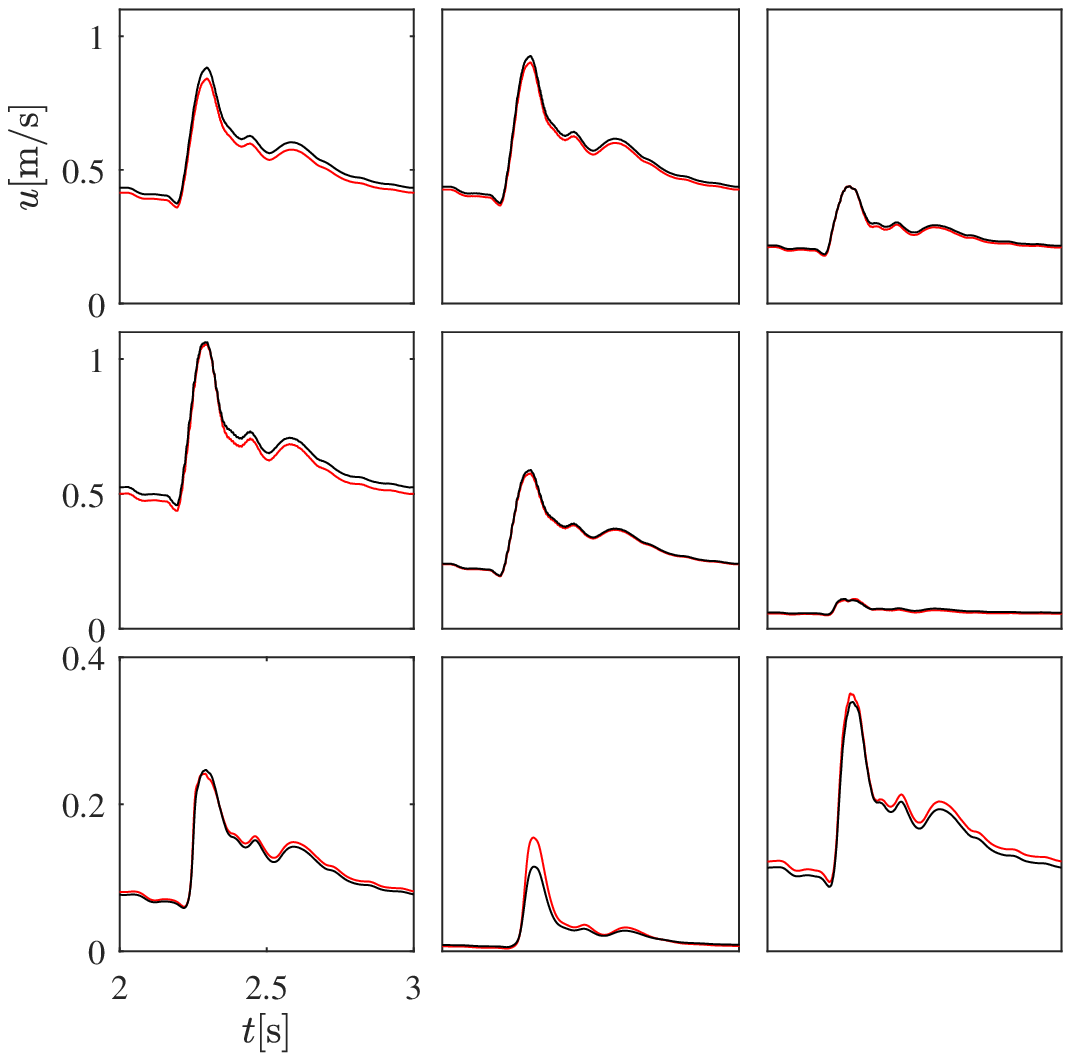}} 
    \end{subfigure}
\caption{Velocity over time in both cases at the 9 monitoring points during the last cardiac cycle. The black and red solid lines denote the velocity magnitude computed using a Newtonian or a non-Newtonian (Cross) model, respectively. From top left to bottom right, results at points P1 to P9 are shown. Note the different vertical scales.}
\label{Fig:VelUn_RuptureNewNonNew}
\end{figure*}
Overall, all simulation results confirm that the flow in the ruptured case exhibits distinct spectral characteristics compared to the unruptured case, especially at the rupture point P6, with a clear frequency peak at approximately 120~Hz before and after the peak systole,
indicating that high-frequency features may play a role in the hemodynamics of aneurysm rupture. 
\subsection{Spectral entropy analysis}
The analysis based on spectral entropy $S_d$ is able to deliver a single, scalar value characterizing the state of an unsteady, 3D flow (as found here). The lowest possible value of $S_d$ (i.e., 0) corresponds to laminar conditions. Previous investigations have shown that transition occurs around $S_d\approx 1$ (see previously cited studies for more detailed values). Here, the values of $S_d$ are compared 1) between ruptured and unruptured aneurysms, and 2) using either Newtonian or non-Newtonian viscosity models. The results are summarized in Table~\ref{table:entropy}. The obtained values of $S_d$ show that the ruptured intracranial aneurysm exhibits significantly higher spectral entropy compared to the unruptured case, corresponding to an unstable flow state with strong fluctuations. The value of $S_d$ for the ruptured case is approximately 10 times larger than for the very stable flow found in the unruptured aneurysm ($S_d \approx 0.1$). This shows that spectral entropy can safely be used to delineate between stable and unstable flow configurations in intracranial aneurysms, as already discussed by Cebral and co-workers~\cite{sforza2016hemodynamics} and Janiga~\cite{janiga2019quantitative}.
\begin{table}[!htb]
\centering
\caption{Values of spectral entropy in the ruptured and unruptured aneurysm for Newtonian or non-Newtonian flow properties.}
\begin{tabular}{llll}
\hline
Category&$S_d$&Flow regime\\
\hline
Unruptured (Newtonian) & 0.10 & Stable\\
Unruptured (non-Newtonian) & 0.12 & Stable\\
Ruptured (Newtonian) & 1.16 & Unstable\\
Ruptured (non-Newtonian) & 1.50 & Unstable\\
\hline
\end{tabular}
\label{table:entropy}
\end{table}
Additionally, it is observed that the spectral entropy of both ruptured and unruptured aneurysms increases by about 20\% when a non-Newtonian model is applied. However, this does not change the predicted flow state (stable for unruptured aneurysm; unstable for ruptured aneurysm). Still, for the ruptured aneurysm, a change of $S_d=1.16$ to $1.5$ corresponds to a noticeable increase in turbulence intensity.
\subsection{Effect of non-Newtonian behavior}
Most of the previous discussion has been based on the assumption of a Newtonian behavior for blood.
Despite the fact that real blood is well-known to be non-Newtonian and is indeed a shear-thinning fluid, the impact of the non-Newtonian properties of blood has been largely overlooked in many past studies; there is still no consensus in the scientific literature regarding the importance of non-Newtonian features, in particular regarding intracranial aneurysms. Blood can be safely assumed as a Newtonian fluid when the shear rates are everywhere greater than 100~$\rm{s^{-1}}$~\cite{fischer2007simulation}. Many published studies argued that the non-Newtonian behavior can be ignored, for instance~\cite{johnston2004non}, while others show dramatic changes when considering non-Newtonian properties~\cite{gijsen1999influence}. The effect of a non-Newtonian behavior on the flow distribution within aneurysms has been recently investigated by Hosseini et al.~\cite{hosseini2022lattice}. Their findings indicate that the selection of a non-Newtonian model has a noticeable impact on the flow field, particularly inside the aneurysm sac and adjacent to its walls, since large parts of the sac are subject to low shear rates. 
\begin{figure}[!ht]
	\centering
   \includegraphics[width=0.9\columnwidth]{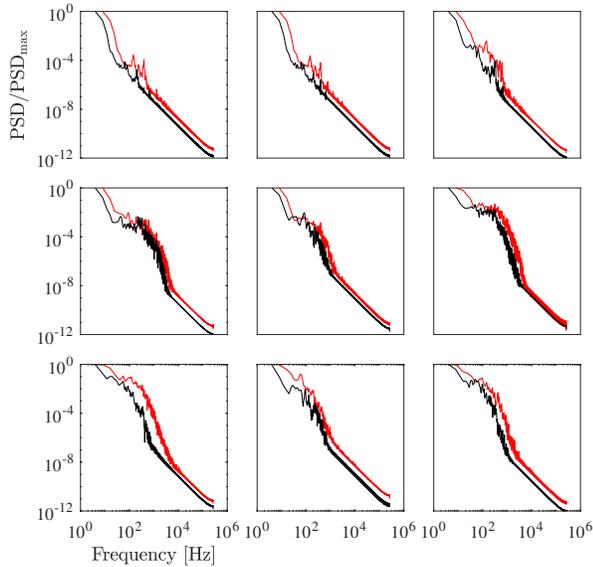}
\caption{Energy spectra of the velocity over frequency in the ruptured case. The black and red solid lines denote the energy spectra computed using a Newtonian or a non-Newtonian (Cross) model, respectively. From top left to bottom right, results at points P1 to P9 are shown.}
\label{Fig:RupNonNewPSD}
\end{figure}
\begin{figure*}[!ht]
	\centering
    \begin{subfigure}[Ruptured case]{\includegraphics[width=0.49\linewidth]{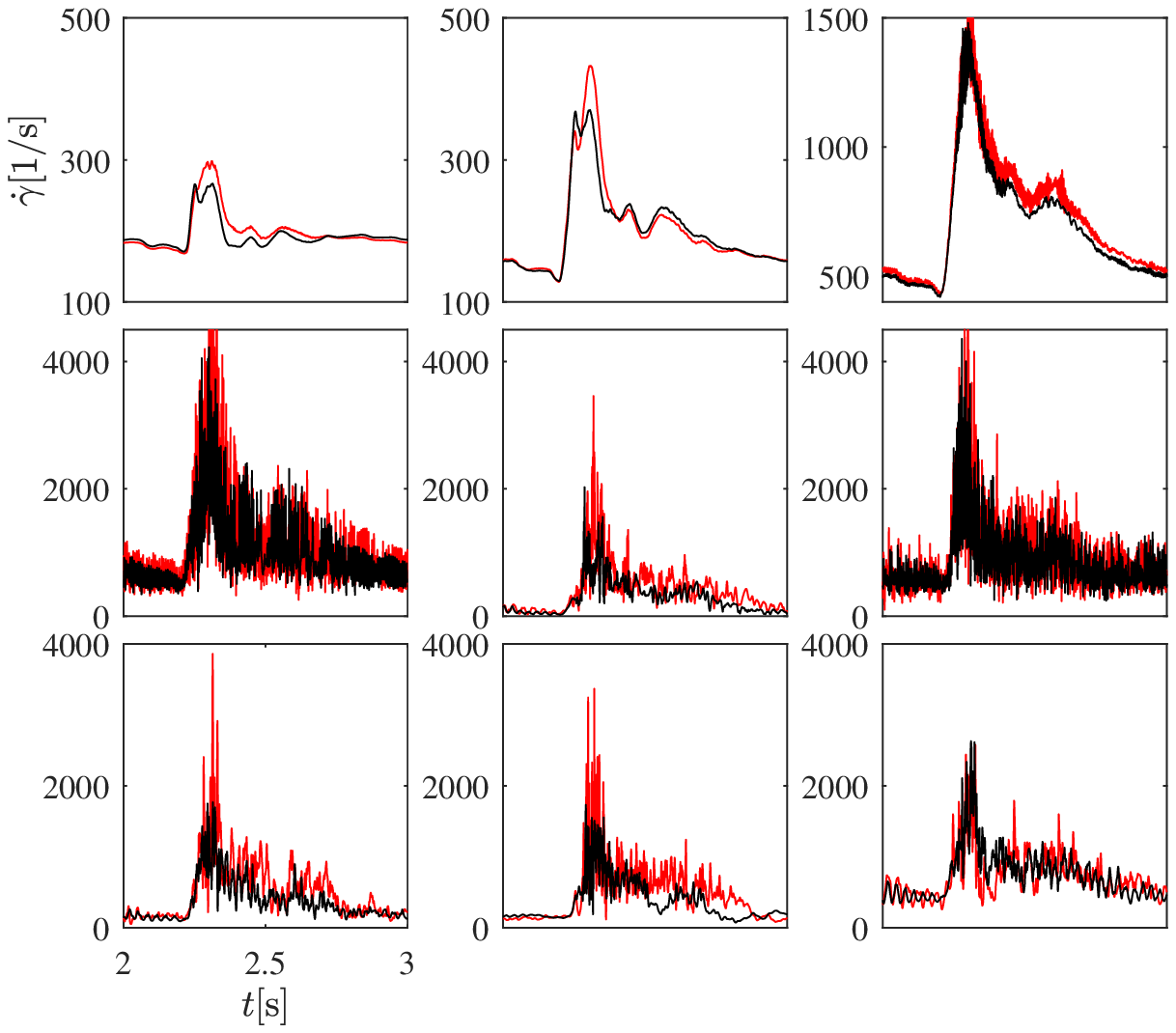}}
    \centering
    \end{subfigure}
    \begin{subfigure}[Unruptured case]{\includegraphics[width=0.49\linewidth]{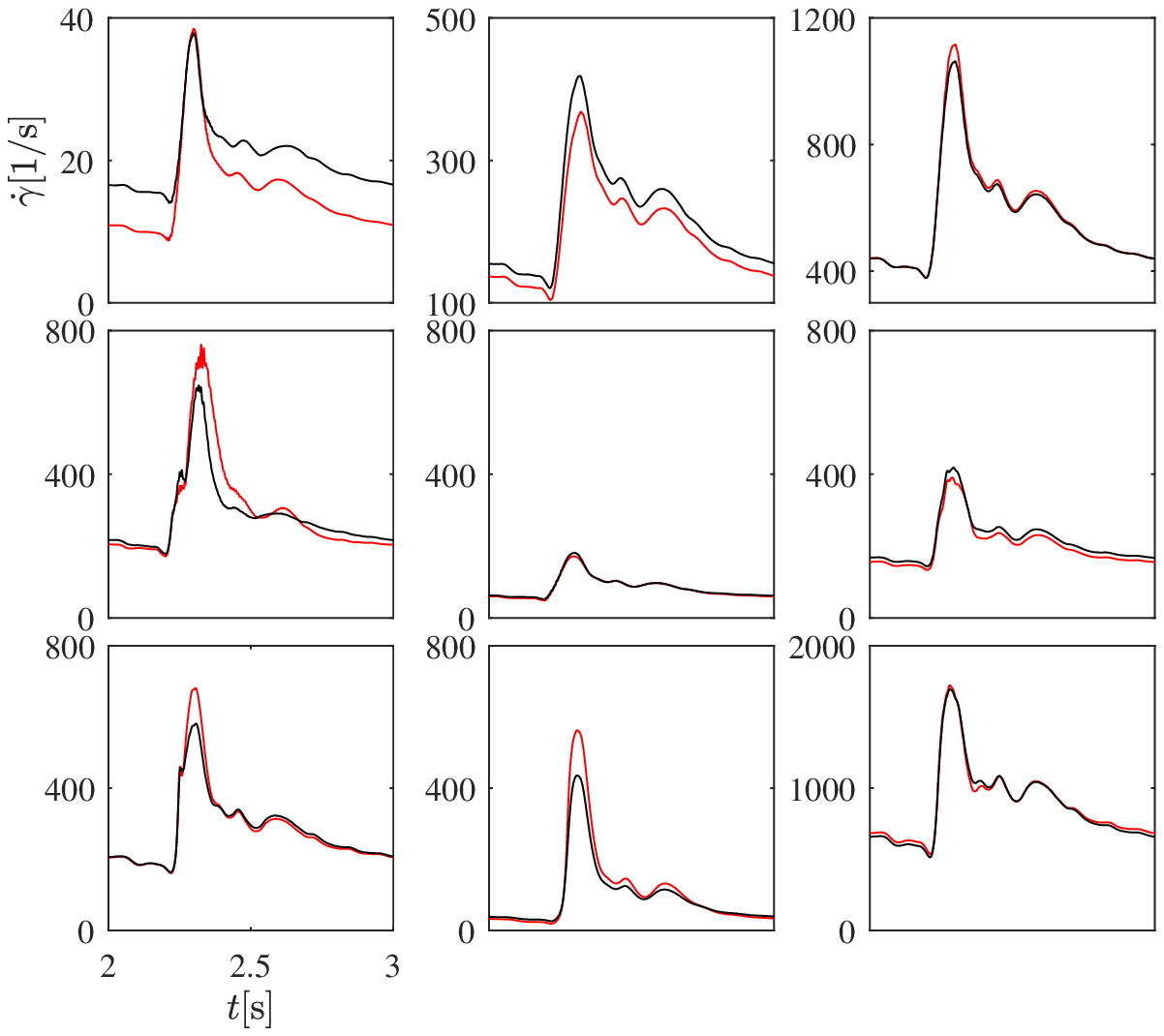}}
    \centering
    \end{subfigure}
\caption{Shear rate over time in both cases at the 9 monitoring points during the last cardiac cycle. The black and red solid lines denote the shear rate computed using a Newtonian or a non-Newtonian (Cross) model, respectively. From top left to bottom right, results at points P1 to P9 are shown. Note the different vertical scales.}
\label{Fig:ShearRateRupture_New_NonNew}
\end{figure*}

In Fig.~\ref{Fig:RuptureUnrup_FlowField}, the streamlines have been qualitatively compared between Newtonian and non-Newtonian cases in both aneurysms, and a clear difference was observed for the ruptured case. Figure~\ref{Fig:VelUn_RuptureNewNonNew} further illustrates the comparison of velocity fluctuations in the ruptured and unruptured case as function of the fluid behavior. The non-Newtonian model has only a small impact on the results for the unruptured case (right part of the figure), as evidenced by the nearly overlapping black and red curves; the only noticeable difference appears at point P8, within the sac. High-frequency fluctuations do not appear, neither assuming a Newtonian nor a non-Newtonian behavior.

The situation is quite different for the ruptured case (left part of Fig.~\ref{Fig:VelUn_RuptureNewNonNew}). The impact of the non-Newtonian model is still negligible in the parent artery (points P1 and P2) and at the neck (P3). However, the velocity fluctuations are much stronger when considering a non-Newtonian behavior, for instance at point P6 (rupture point). At point P8, even the mean velocity profile is completely different when taking into account non-Newtonian effects. This explains why the streamlines have been found different as well in Fig.~\ref{Fig:RuptureUnrup_FlowField}. Figure~\ref{Fig:RupNonNewPSD} presents the energy spectra of the velocity in the ruptured case. It can be observed that the non-Newtonian model exhibits much higher energy levels compared to the Newtonian model. These results suggest that a non-Newtonian fluid behavior may enhance flow instability. \\
\indent Additionally, the local shear rate at the different monitoring points is shown in Fig.~\ref{Fig:ShearRateRupture_New_NonNew} to investigate the possible impact of the different viscosity models. In the unruptured case, the evolution of the shear rates is similar considering a Newtonian or a non-Newtonian model, with maximum relative differences around peak systole below 25\%; high-frequency fluctuations are never observed. In the unruptured case, the shear rates are quite small and stay below 400~$\rm{s^{-1}}$ for most of the points.\\
\indent In the ruptured case, this is only true within the cerebral artery before the aneurysm (points P1$\&$P2). The fluctuation in shear rate is still small at the bifurcation point (P3), while very large fluctuations are observed at all further points P4 to P9. The peak shear rate even exceeds 4000~$\rm{s^{-1}}$ at peak systole at the outlet P4 (that with the largest flow-rate, as discussed previously) and at the rupture point P6. This finding confirms the presence of significant velocity variations at these two points. 

In all cases, the non-Newtonian simulations exhibit larger amplitudes of shear rate fluctuations. This is particularly true at peak systole; at P5, P7, and P8, the peak shear rate reaches almost 4000~$\rm{s^{-1}}$ with the non-Newtonian model, twice as high as when assuming a Newtonian behavior. Unexpectedly, the rupture point P6 is not strongly affected, and the shear rates are there very similar for the Newtonian and the non-Newtonian model.
\section{Conclusions and discussion\label{sec:discussion}}
This comparative study provides valuable insights regarding hemodynamic characteristics in a ruptured and an unruptured, patient-specific intracranial aneurysm. These findings suggest that a high level of spatial and temporal resolution is required to investigate high-frequency fluctuations possibly occurring in such configurations; using a coarser grid and a large time step leads to increased numerical dissipation and damps artificially the onset of flow instability.\\
\indent Furthermore, these investigations also demonstrate that non-Newtonian models may have a very large impact on shear rate at specific locations, particularly around peak systole. This is even more important in the ruptured aneurysm. However, the exact rupture point is unexpectedly almost not affected by the shear rate -- so that further studies involving additional configurations will be needed to clarify the picture. Still, it cannot be assumed that the shear rate will exceed 100~$\rm{s^{-1}}$ at all locations and at all times, so that using a suitable non-Newtonian model is recommended to be on the safe side; this might be particularly important when assessing flow instability in patient-specific aneurysms.\\
\indent Several limitations should be mentioned in connection with the present study. The first one is obviously the very limited number of configurations considered -- only one ruptured and one unruptured case. It is well known that aneurysm size, shape, location, the ratio of size and neck\ldots will affect hemodynamics and aneurysm rupture. A much larger number of patient-specific configurations must be considered in future studies. Secondly, all vessels have been assumed to be rigid. Clearly, the obtained findings might be modified when taking into account fluid-structure interactions; however, this would only be possible when having a detailed knowledge of wall resistance and structure, which is extremely challenging~\cite{voss2018fluid,voss2016fluid}.\\
\indent Finally, it is worth noting that the interplay between hemodynamic factors and biological processes regarding aneurysm formation and rupture remain poorly understood. Best-practice rules are needed for the assessment of aneurysm hemodynamics using numerical simulations, in order to support more robust and reliable comparisons among studies~\cite{Janiga2015Challenge,Berg2019Review,berg2019multiple}. 
Ultimately, this would open the door for an efficient support of clinicians, leading to personalized treatment options.
\section*{Acknowledgment}
F.H. would like to acknowledge the financial support of the China Scholarship Council (grant number 201908080236). The authors also gratefully acknowledge the Gauss centre for providing computation time under grant "pn73ta" on the GCS supercomputer superMUC-NG at Leibniz supercomputing center, Munich, Germany. The authors further thank Dr. Samuel Voß and Dr. Oliver Beuing for their assistance in acquiring the patient-specific models.
\nocite{*}

\section*{AUTHOR DECLARATIONS}

\noindent\textbf{Conflict of Interest}\\

\noindent The authors have no conflicts to disclose.

\section*{Data Availability Statement}

\noindent The data that support the findings of this study are available from the corresponding author upon reasonable request.

\section*{references}
\bibliography{refs}
\end{document}